\documentclass[11pt,fleqn]{article} 
\usepackage{psfig}
\title{Possible Enhanced Flux of Glassy Solid \\ Helium in Cylindrical Corrugated Nanopores}  %
\author{Kwang Hua Chu and Zhu Guang Hua} 
\date{Department of Physics, Wulumuqi Toudiban (Transfer Centre), Xinjiang 830000, China}
\begin{document}           
\doublerulesep=6.5mm        
\baselineskip=6.5mm
\oddsidemargin-1mm         
\maketitle
\begin{abstract}
By using the glassy (helium-)fluid model and boundary perturbation
method, we can  obtain the velocity fields (as well as the flow
rate; up to the second order) inside the wavy-rough cylindrical
nanopores which are of the same size as those samples prepared by
Kim and Chan as well as Day {\it et al.} Our results show that the
velocities measured in porous Vycor samples could be reproduced by
carefully selecting relevant parameters but those in glass
capillaries are difficult to obtain. \newline

\noindent Keywords : Supersolidity, boundary perturbation approach
\newline

\noindent PACS : 67.80.-s, 67.40.Hf
\end{abstract}
\bibliographystyle{plain}
\section{Introduction}
A supersolid (SS) is a solid that possesses an element of
superfluidity$^1$. The existence of SS $^4$He was first suggested 67
years ago$^2$, while possible underlying mechanisms were discussed
30 years later$^{3-4}$. The subsequent crucial step is the detection
of solid $^4$He acquiring nonclassical rotational inertia (NCRI)$^5$.
NCRI is indicated by a drop in the resonant period of the
oscillator below the transition temperature. Recently the
experimental evidence for SS was reported by Kim and Chan  in a
series of torsional oscillator experiments on bulk solid helium$^6$
where it was demonstrated that the decrement in the period is
proportional to the SS fraction in the limit of low oscillation
speed.
\newline
Before the discovery of SS, there already were many researches$^{7-12}$
 about its essence and existence. However, even after Kim and
Chan announced the experimental evidence of SS, Day {\it et al.}
conducted similar experiments$^{13}$ and showed no indication of a
mass redistribution in the Vycor that could mimic SS decoupling and
put an upper limit of about 0.003 $\mu$m/s on any pressure-induced
SS flow in the pores of Vycor. Later on Day and Beamish's
measurements$^{14}$ showed no indication of low temperature flow,
placing stringent restrictions on SS flow in response to a pressure
difference. The average flow speed at low temperatures was less than
1.2 $\times$ 10$^{-14}$ m/s, corresponding to a SS velocity at least
7 orders of magnitude smaller than the critical velocities inferred
from the torsional oscillator measurements$^{15-16}$. Thus it was
claimed that the origin or exact mechanism of free-flowing SS
although confirmed but not yet resolved$^{17-22}$. Meanwhile, as
evidenced by recent experiment (by Grigorev {\it et al.})$^{23}$ where
the temperature dependence of pressure in solid $^4$He grown by the
capillary blocking technique was measured; at temperatures below
0.3K (where the supersolidity was observed) they found the glassy
$\propto$ T$^2$ contribution to pressure. It was then proposed by
Andreev$^{24}$ and Rittner and Reppy$^{25}$ the observed supersolidity
might be in a glassy solid (helium) state. Both idea could also be
inspired from the annealing effect of supersolidity reported before$^{19}$.
\newline
In this short paper, we offer an explanation for the argues between
Day {\it et al.}$^{13-14}$ and Kim and Chan$^{15-16}$ by treating
the nanopores which were produced in porous Vycor$^{15}$ or porous
gold samples$^{16}$ to be a kind of (cylindrical) wavy-rough
nanotubes. The transport of glassy SS helium$^{21}$ or quantum
glass$^{24}$ inside nanopores which have presumed wavy-rough wall
will be investigated here. We adopt the verified model initiated by
Cagle and Eyring$^{25}$ which was used to study the annealing of
glass. To obtain the law of annealing of glass for explaining the
too rapid annealing at the earliest time, because the relaxation at
the beginning was steeper than could be explained by the bimolecular
law, Cagle and Eyring tried a hyperbolic sine law between the shear
(strain) rate : and (large) shear stress : $\tau$ and obtained the
close agreement with experimental data. This model has sound
physical foundation from the thermal activation process$^{26}$
(Eyring$^{26}$ already considered a kind of (quantum) tunneling
which relates to the matter rearranging by surmounting a potential
energy barrier) and thus it might also resolve the concern raised by
Anderson$^{22}$ for the thermal noises to the superflow of vortex
liquid. With this model we can associate the (glassy) fluid with the
momentum transfer between neighboring atomic clusters on the
microscopic scale and reveals the atomic interaction in the
relaxation of  flow with dissipation (the momentum transfer depends
on the activation shear volume, which is associated with the center
distance between atoms and is proportional to $RT/\tau_0$ ($R$ is
the gas constant, $T$ temperature in Kelvin, and $\tau_0$ a constant
with the dimension of stress).
\newline To consider the more realistic but complicated  boundary
conditions in the wall of nanotubes, however, we will use the
boundary perturbation technique$^{27}$ to handle the presumed
wavy-roughness along the wall of nanotubes. The relevant boundary
conditions along the wavy-rough surface will be prescribed below.
The contents are organized into three parts : Introduction is
firstly presented and then we describe our physical assumptions and
the mathematical model. After that we illustrate our results and
possible comparison with previous rather scattered measurements. In
fact, the qualitative agreement is rather good!
\section{Physical and Mathematical Formulations}
We shall consider a steady transport of the glassy SS helium in a
wavy-rough nanotube of $a$ (mean-averaged radius) and the outer
wall being a fixed wavy-rough surface, $r=a+\epsilon \sin(k
\theta)$, where $\epsilon$ is the amplitude of the (wavy)
roughness, and the wave number : $k=2\pi /L $. Firstly, this fluid$^{
25-26}$ can be expressed as
 $\dot{\gamma}=\dot{\gamma}_0  \sinh(\tau/\tau_0)$,
where $\dot{\gamma}$ is the shear rate, $\tau$ is the shear stress,
and $\dot{\gamma}_0$ is a function of temperature with the dimension
of the shear rate. In fact, the force balance gives the shear stress
at a radius $r$ as $\tau=-(r \,dp/dz)/2$. $dp/dz$ is the pressure
gradient along the flow (or tube-axis : $z$-axis) direction.\newline
Introducing
$\chi = -(a/2\tau_0) dp/dz$
then we have
 $\dot{\gamma}= \dot{\gamma}_0  \sinh ({\chi r}/{a})$.
As $\dot{\gamma}=- du/dr$ ($u$ is the velocity of the fluid flow in
the longitudinal ($z$-)direction of the nanotube), after
integration, we obtain
\begin{equation}
 u=u_s +\frac{\dot{\gamma}_0 a}{\chi} [\cosh \chi - \cosh (\frac{\chi r}{a})],
\end{equation}
here, $u_s$ is the velocity over the surface of the nanotube, which
is determined by the boundary condition.  We noticed that Thompson
and Troian$^{28}$ proposed a general boundary condition for trasnport
over a solid surface as
\begin{equation}
 \Delta u=L_s^0 \dot{\gamma}
 (1-\frac{\dot{\gamma}}{\dot{\gamma}_c})^{-1/2},
\end{equation}
where  $\Delta u$ is the velocity jump over the solid surface,
$L_s^0$ is a constant slip length, $\dot{\gamma}_c$ is the critical
shear rate at which the slip length diverges. The value of
$\dot{\gamma}_c$ is a function of the corrugation of interfacial
energy.  \newline With the boundary condition from Thompson and
Troian$^{28}$,  we can derive the velocity field and volume flow rate
along the wavy-rough nanotube below using the verified boundary
perturbation technique$^{27}$. The wavy boundary is prescribed as
$r=a+\epsilon \sin(k\theta)$ and the presumed steady transport is
along the $z$-direction (nanotube-axis direction). \newline Along
the boundary, we have
 $\dot{\gamma}=(d u)/(d n)|_{{\mbox{\small on surface}}}$.
Here, $n$ means the  normal. Let $u$ be expanded in $\epsilon$ :
 $u= u_0 +\epsilon u_1 + \epsilon^2 u_2 + \cdots$,
and on the boundary, we expand $u(r_0+\epsilon dr,
\theta(=\theta_0))$ into
\begin{displaymath}
u(r,\theta) |_{(r_0+\epsilon dr ,\theta_0)}
=u(r_0,\theta)+\epsilon [dr \,u_r (r_0,\theta)]+ \epsilon^2
[\frac{dr^2}{2} u_{rr}(r_0,\theta)]+\cdots=
\end{displaymath}
\begin{equation}
 \hspace*{12mm} \{u_{slip} +\frac{\dot{\gamma} a}{\chi} [\cosh \chi - \cosh (\frac{\chi
 r}{a})]\}|_{{\mbox{\small on surface}}},
\end{equation}
where
\begin{equation}
 u_{slip}|_{{\mbox{\small on surface}}}=L_S^0 \dot{\gamma} [(1-\frac{\dot{\gamma}}{\dot{\gamma}_c})^{-1/2}]
 |_{{\mbox{\small on surface}}}, \hspace*{2mm}
 u_{{slip}_0}= L_S^0 \dot{\gamma}_0 [\sinh\chi(1-\frac{\sinh\chi}{
 \dot{\gamma}_c/\dot{\gamma}_0})^{-1/2}].
\end{equation}
Now, on the outer wall$^{27}$
\begin{displaymath}
 \frac{du}{dn}=\nabla u \frac{\nabla (r-a-\epsilon
\sin(k\theta))}{| \nabla (r-a-\epsilon \sin(k\theta)) |}
=[1+\epsilon^2 \frac{k^2}{r^2}  \cos^2 (k\theta)]^{-\frac{1}{2}}
[u_r |_{(a+\epsilon dr,\theta)} -
\end{displaymath}
\begin{displaymath}  
 \hspace*{12mm} \epsilon \frac{k}{r^2}
\cos(k\theta) u_{\theta} |_{(a+\epsilon dr,\theta)} ]=u_{0_r}|_{a}
+\epsilon [u_{1_r}|_{a} +u_{0_{rr}}|_{a} \sin(k\theta)-
\end{displaymath}
\begin{displaymath}
  \hspace*{12mm}  \frac{k}{r^2} u_{0_{\theta}}|_{a} \cos(k\theta)]+\epsilon^2 [-\frac{1}{2} \frac{k^2}{r^2} \cos^2
(k\theta) u_{0_r}|_{a} + u_{2_r}|_{a} + u_{1_{rr}}|_{a} \sin(k\theta)+ 
\end{displaymath}
\begin{equation}
   \hspace*{12mm} \frac{1}{2} u_{0_{rrr}}|_{a} \sin^2 (k\theta) -\frac{k}{r^2}
\cos(k\theta) (u_{1_{\theta}}|_{a} + u_{0_{\theta r}}|_{a}
\sin(k\theta) )] + O(\epsilon^3 ) .
\end{equation}
Considering $L_s^0 \sim a \gg \epsilon$ case, we also presume
$\sinh\chi \ll \dot{\gamma}_c/\dot{\gamma_0}$.
With equations (1) and (5), using the definition of $\dot{\gamma}$,
we can derive the velocity field ($u$) up to the second order. The
key point is to firstly obtain the slip velocity along the boundary
or surface.
After lengthy mathematical manipulations, we obtain %
the velocity fields (up to the second order) and then we can
integrate them with respect to the cross-section to get the volume
flow rate ($Q$, also up to the second order here) :
\begin{equation}
  Q=\int_0^{\theta_p} \int_0^{a+\epsilon \sin(k\theta)} u(r,\theta) r
 dr d\theta =Q_{smooth} +\epsilon\,Q_{p_0}+\epsilon^2\,Q_{p_2}.
\end{equation}
Here, $\theta_p$ could be $2\pi$ or $2\pi/k$ depending on the
specific requirement for dimensionless consideration.
\section{Results and Discussions}
We firstly compare the velocity fields between the smooth and
wavy-rough nanotubes in Fig. 1 for the mean radius $a=3.5$ nm (cf.
Refs. 15 and 20). The wave number $k$ is fixed to be 10. We select
the amplitude of wavy-roughness $\epsilon$ to be  0.02 and 0.04
$a$ (to check the geometry effect which is valid for small
$\epsilon$ here due to the perturbation approach). We try
$\dot{\gamma}_0 =10000$ s$^{-1}$,
$\dot{\gamma}_0/\dot{\gamma}_c$=0.1 (cf. Ref. 29). We then set
$L_s^0= 0.5 a$. The x-axis ($\chi/a$) shown in Figure 1 is the
ratio of forcing (along the $z$-axis direction) per unit volume
and (shear) stress. The y-axis is for the total velocity ($u$), up
to the second order. Note that the real range of the (referenced)
shear rate : $\dot{\gamma}_0$ will depend on the specific material
chosen as well as the experimental procedure$^{29}$. We can
observe the wavy-roughness ($\epsilon$) will significantly enhance
the flow rate$^{30}$ due to larger surface-to-volume ratio and
{\it slip} effect (along the wall)$^{31}$. There is net flow rate
even $\chi/a$ is zero (without forcing!)$^{30}$ (it could be a
persistent current or  a spontaneous flow without a pressure
difference and the velocity of this supeflow is proportional to
the small amplitude of wavy-roughness, the (referenced) shear rate
and the slip length). Our results agree with those of Ref. 21
(measured higher SS fraction) for larger disordered or glassy
helium cases (having much larger roughness than that of Refs. 15
and 16).\newline Once $\chi \sim 0.1$, the velocity for the smooth
nanopore will be around 10 $\mu$m/s which was reported in Ref. 15.
The upper limit of $u$ set in Ref. 13 by Day {\it et al.} could be
due to $\chi$ less than $0.0001$ (presumed their sample of smooth
nanopores, too)! We remind the readers that the geometry effect
for the enhanced flow rate (for small wavy-roughness and smaller
forcing) is clear as shown in this figure (once $\epsilon$ is
increasing) but we should keep in mind that the perturbation
approach might limit $\epsilon$ to be less than 0.1 $a$.
\newline
To check the larger pore effect ($a=245$ nm in Ref. 16), we calculate
the total velocity for $a$=245 nm case while keeping all other
parameters almost the same. The trend (differences between the
smooth and rough nanotubes) shown in Fig. 2 is almost the same as
that in Figure 1. We can observe that, under the same selected
parameters, for larger smooth nanopores, the velocity should be
much larger than that in smaller smooth nanopores (at least 2
order of magnitudes larger). However, as reported in Ref. 16 for
porous gold cases, the velocity is almost the same order of
magnitude as that in Ref. 15. There might be agreement between our
results and those in Ref. 16 for $a=245$ nm case if $\dot{\gamma}_0
=10000$ s$^{-1}$ is lowered to less than $100$ s$^{-1}$! As the
nanopore size in Ref. 14  (glass capillaries) is unknown, thus to
obtain the rather small velocity ($\le 1.2 \times 10^{-6} \mu$m/s)
reported therein $\chi$ should be less than $10^{-8}$. Note that
those lowest curves illustared in Figs. 1 and 2 were obtained
using the Navier-slip boundary conditions (e.g., cf. Ref. 32).
\newline In brief summary, we have theoretically obtained the velocity
(up to the second order) inside the wavy-rough cylindrical
nanopores by using the glassy (solid) helium-fluid model and
boundary perturbation method. Our results show that the calculated
velocity for smaller (presumed) smooth nanopores (radius $\sim
3.5$ nm) could be of the same order of magnitude as those in Ref. 15
after carefully selected parameters. Those measurements reported
in Ref. 14, however, as the size of nanopores are unknown, are
difficult to be reporduced using our approach!
We shall investigate more complicated issues$^{33}$  in the future.

\newpage

\psfig{file=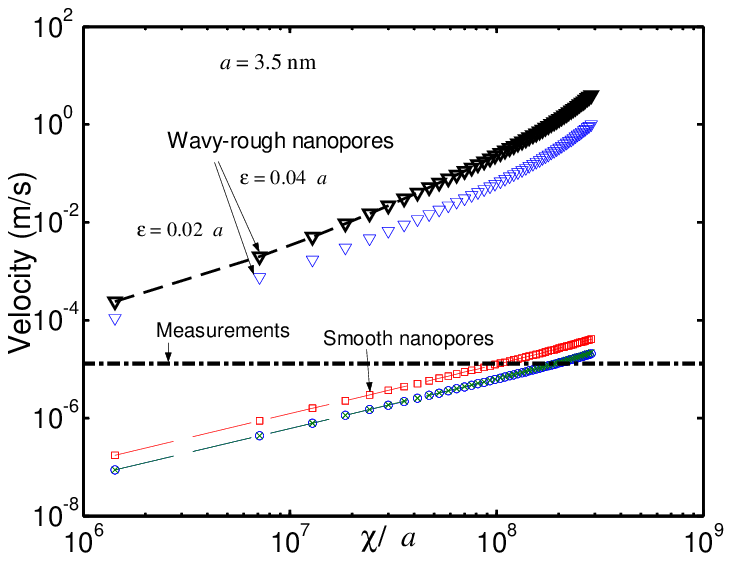,bbllx=0.0cm,bblly=15cm,bburx=8cm,bbury=25cm,rheight=8cm,rwidth=8cm,clip=}

\begin{figure}[h]
\hspace*{8mm} Fig. 1 \hspace*{1mm} Calculated velocity fields for
different $\chi/a$ (forcing (along the $z$-axis direction)
\newline \hspace*{8mm} per unit volume and referenced shear
stress). $a=3.5$ nm. $\epsilon$ (=0.02 and 0.04 $a$ here) is the
\newline \hspace*{8mm} amplitude of wavy roughness. The wave number
of roughness ($k$) is $10$ here. Kim and \newline \hspace*{8mm}
Chan's data$^{15}$ could be reproduced once $\chi$ is around 0.1
and the slip length ($L_s^0$) is 0.5 $a$.
\newline \hspace*{8mm} The lowest curves are only due to boundary effects
(i.e., boundary slip or contributions \newline \hspace*{8mm} from
the slip length term in equation
 (7)) for smooth nanopores. \newline \hspace*{8mm} As the
 experimental data were quite scattered and we only put the roughly
 averaged value \newline \hspace*{8mm} here for comparison.
\end{figure}

\newpage

\psfig{file=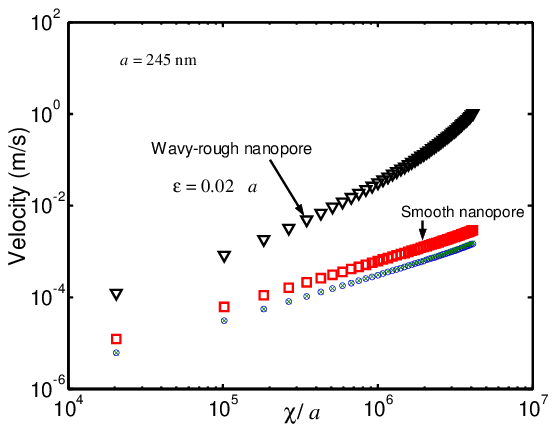,bbllx=-0.5cm,bblly=23cm,bburx=8cm,bbury=29cm,rheight=6cm,rwidth=6cm,clip=}

\begin{figure}[h]
\hspace*{10mm} Fig. 2 \hspace*{1mm} Calculated velocity fields for
different $\chi/a$ (forcing (along the $z$-axis direction)
\newline \hspace*{10mm} per unit volume and referenced shear
stress). $a=245$ nm. $\epsilon$ (=0.02 $a$ here) is the
\newline \hspace*{10mm} amplitude of wavy roughness. The wave
number of roughness ($k$) is $10$ here. Kim and \newline
\hspace*{10mm} Chan's data$^{16}$ could be reproduced once
$\dot{\gamma}_0$ (the shear rate) is lowered down to $100$
s$^{-1}$.
\newline \hspace*{10mm} The lowest curves are only due to boundary effects
(i.e., boundary slip or contributions \newline \hspace*{10mm} from
the slip length term in equation
 (7)).
\end{figure}
%
\end{document}